\documentclass[]{JHEP3}
\usepackage{epsf,graphicx}
\usepackage{latexsym}
\usepackage[latin1]{inputenc}
\usepackage[T1]{fontenc}
\title{Non-BPS Brane Cosmology}

\author{Ph.Brax \\ Service de Physique Th\'eorique,
CEA Saclay, 91191 Gif-sur-Yvette, France \\ E-mail: \email{
brax@spht.saclay.cea.fr}}

\author{D.A.Steer \\ Laboratoire de
Physique Th\'eorique, B\^at. 210, Universit\'e Paris XI, \\ 91405
Orsay Cedex, France \\E-mail: \email{steer@th.u-psud.fr}}

\abstract{ We study cosmology on a BPS D3-brane evolving in the
10D SUGRA background describing a non-BPS brane. Initially the BPS
brane is taken to be a probe whose dynamics we determine in the
non-compact non-BPS background.  The cosmology observed on the
brane is of the FRW type with a scale factor $S(\tau)$. In this
mirage cosmology approach, there is no self-gravity on the brane
which cannot inflate.  Self-gravity is then included by
compactifying the background space-time. The low energy effective
theory below the compactification scale is shown to be bi-metric,
with matter coupling to a different metric than the geometrically
induced metric on the brane. The  geometrical scale factor on the
brane is now $S(\tau) a(\tau)$ where $a(\tau)$ arises from brane
self-gravity. In this non-BPS scenario the brane generically
inflates. We study the resulting inflationary scenario  taking
into account the fact that the non-BPS brane eventually decays on
a time-scale much larger than the typical inflationary time-scale.
After the decay, the theory ceases to be bi-metric and COBE
normalization is used to estimate the string scale which is found
to be of order $10^{14}$ GeV.}


\preprint{ORSAY-LPT-02-33, T/02-048}

\begin{document}

\section{Introduction}
Since the advent of branes in string theory, brane models have
been used to improve our understanding of physics beyond the
standard model. Brane constructions have proved to be
particularly successful in providing valuable insights for  such
arduous puzzles as the cosmological constant problem and the
hierarchy problem. In their most common guise, models of this type
are five-dimensional following the work of Randall-Sundrum
\cite{RSI,RSII} and the compactification of the heterotic M-Theory
on a Calabi-Yau manifold \cite{HW}.  Our universe is then one of
the boundaries of the fifth dimension.

Another approach has been followed starting from ten dimensional
string theory compactified on a large volume manifold. These
models allow one to lower the string scales in such a way that
stringy effects may become observable at accelerators. As opposed
to the five dimensional approach the geometry of space-time is
flat \cite{AR,AN}.

One of the puzzles of high energy physics is supersymmetry
breaking. A particularly useful way of breaking supersymmetry has
been studied with branes, i.e.\ configurations of branes and
anti-branes do not preserve any supersymmetry because of the
incompatibility of their respective  BPS conditions. The lack of
supersymmetry implies the existence of a tachyon in the open
string spectrum, signaling an instability
\cite{BI,BII,KI,SI,SII,SIII,SIV,KII,KR,T,SV}. Indeed
brane-antibrane configurations are not static due the existence of
a long-range force between them. Such a force can be evaluated in
the large separation regime, and hence at low energy the brane
separation can be viewed as a field with a non-trivial potential.
This setting has been used to model inflation in the primordial
universe \cite{MI,DI,DII,BUI,GI,AI,HI,SH,HE,KY,BUII,BL,D,J,MA,O}.
As long as the brane and anti-brane are far apart one can trust
the supergravity approximation. When the branes are nearly
colliding, the open string description is more appropriate. In
that case, the Sen conjecture  \cite{SI}ascertains that the open string
tachyon condensates leaving behind a supersymmetric configuration.
For brane-antibrane, the result of the annihilation can either be
flat empty space or higher codimension supersymmetric branes\cite{SIV}. This
idea has been used to justify the end of inflation and the
occurrence of reheating.

One may also be interested in non-BPS branes per se, i.e. branes
which breaks all supersymmetries. These objects can be constructed
in string theory at small string coupling by considering a
brane-antibrane configuration after modding out $(-1)^{F_L}$, the
left fermion number\cite{SIII}. This gives rise to non-BPS branes
with a real tachyon in the open string spectrum. As
brane-antibranes, they are meant to decay via open string tachyon
condensation. Nevertheless one may be interested in describing the
non-BPS branes from the supergravity point of view, i.e.\ the
classical configuration of the closed string zero modes which
results from the distortion of space-time by the presence of a
non-BPS brane. Such a programme has been carried out in
\cite{BMO}. The tachyon condensation becomes a path in the
parameter space interpolating between the non-BPS branes and the
BPS branes.

In this paper we consider the low energy physics viewed on a probe
brane in the background of a non-BPS brane. The probe is taken to
be far enough away from the non-BPS brane that the supergravity
approximation can be trusted, and analyse the dynamics before the
decay of the non-BPS brane. We consider two situations. In section
2, we study the case of a $D3$ brane probing the geometry of a
non-BPS brane in non-compact ten dimensional space.  In that case
there is no self-gravity on the brane, though there is an
effective cosmology. We show that despite the attraction between
the branes, inflation never occurs. Then in section 3 we
compactify the model on a large volume six-manifold that we do not
need to specify. We will consider that the moduli resulting from the
compactification have been stabilized by an unknowm mechanism.
In particular the overall volume of compactification will taken to be
a free parameter of the model.  As a
result of self-gravity, the effective four dimensional Planck
mass is finite and  the brane undergoes a phase of inflation. After
the decay of the non-BPS brane, the probe brane reheats. Inflation
does not require any fine-tuning of the parameters of the non-BPS
brane, i.e. it takes place whether the non-BPS brane is at the
beginning or end of its decay. We stress that the only fine-tuning
comes from the initial position of the probe which determines the
number of e-folds of inflation.

In section 2, where the geometry is non-compact, the brane
`mirage' cosmology arises solely due to the brane motion
\cite{SP,KK,KIRI,KIRII}.  The result of compactification, in
section 3, leads us to consider a 4d effective action similar to
that of \cite{MI,DI,DII,BUI,GI,AI,HI,SH,HE,KY,BUII,BL,D,J,MA,O}
which includes self-gravity. However, as opposed to those
references, our theory on the brane turns out to be bi-metric.
The setup we envisage is also different to that of \cite{BUI,BUII}
who suppose that inflation ends in a $\bar{D3}-D3$ collision which
will generically produce  branes of higher codimension. Here the
non-BPS brane eventually decays leaving only our $D3$-brane and
radiation. The scenario again differs from that of \cite{GI,J} who
considered non-parallel branes, and of reference \cite{AI} in
which inflation occurs during a $\bar{D5}-D5$ annihilation to
produce three-branes.

\section{Non-BPS background metric}

In the supergravity approximation, where only massless closed
string modes are taken into account, the background generated by a
non-BPS $p$-brane at the origin was constructed in \cite{BMO,ZZ}.
This background metric, in the Einstein frame, is given by
\begin{eqnarray}
ds_E^2 &=& e^{2 A(r)}(-dt^2 + dx_i dx^{i}) + e^{2 B(r)} dr^2 + r^2
e^{2 B(r)} d\Omega_{(8-p)}^2 \nonumber
\\
&\equiv& \bar{G}_{AB} dx^{A} dx^{B} \label{metricE}
\end{eqnarray}
where $i=1,\ldots,p$ and capital latin indices run over the 10
spacetime indices ($A,B = 0,\ldots,9$). Here we focus on
D3-branes:  for $p=3$ the functions appearing in (\ref{metricE})
are given by \cite{BMO}
\begin{equation}
A(r) = -\frac{1}{4} \ln Y , \qquad B(r)  =  \frac{1}{4} \ln[f_+
f_-] - A(r)
\end{equation}
where
\begin{eqnarray}
Y & = &  \cosh( k \,  h(r) )  - c_2\, \sinh( k \, h(r) ) \nonumber
\\
f_{\pm} &=& 1 \pm \left(\frac{r_0}{r} \right)^4 \nonumber
\\
h & = & \ln[\frac{f_-}{f_+}] \nonumber
\\
k &=&  \sqrt{{5\over 2} - c_1^2} . \nonumber
\end{eqnarray}
The corresponding dilaton $\phi(r)$ and 4-form
$C^{(4)}=e^{\Lambda(r)} dx^0 \wedge dx^1 \wedge dx^2 \wedge dx^3$
are given by
\begin{eqnarray}
\phi & = & c_1 h(r) \nonumber
\\
 e^{\Lambda}  &=&  - \tilde{\eta} (c_2^2-1)^{1/2} \,
{\sinh (k \, h(r) ) \over Y }
\label{lambda}
\end{eqnarray}
where $\tilde{\eta}$ is background non-BPS ``brane'' or
``antibrane'' charge.

Notice from (\ref{lambda}) that $|c_2| \geq 1$, and also that
$c_1^2 \leq 5/2$ since $k$ must be real (see below).  In the
following we take $r_0 > 0$, and the resulting naked singularity
at $r=r_0$ reflects the lack of supersymmetry of the system.
(Hence $r \geq r_0$; for a discussion of the case in which $r_0 <
0$, see \cite{BMO}.)  Indeed, recall that stable supergravity
configurations are specified by two parameters corresponding to
the ADM mass, $M_{ADM}$, and the charge, $Q$, as required by
Birkhoff's theorem. Here, though, a third parameter $c_1$ appears
due to the non-trivial tachyon v.e.v.~$T$.  As a result Birkhoff's
theorem must break down and this leads to the naked singularity.
More specifically, for metric (\ref{metricE})
\begin{eqnarray}
M_{ADM}&=&2c_2k N_3 r_0^4 , \nonumber
\\
Q &=& 2\eta N_3 r_0^4 \sqrt{c_2^2-1},
 \label{charge}
\end{eqnarray}
where $N_3= 5\omega_5 V_3/4\kappa_{10}^2$, $\omega_5=\pi^3$, and
for convenience we have wrapped the spatial world-volume
directions on a torus of volume $V_3$ \cite{BMO}.  Notice from
(\ref{charge})
%, (\ref{lambda})
that the parameter $c_2$ determines the charge of the background, i.e.
when $c_2^2=1$ the background is neutral.  In sections 3 and 4 we
will mainly focus on that case. Also notice that the BPS relation
$M_{ADM}=Q$ is generally violated.

The physical interpretation of $c_1$ is that it is related to $T$
\cite{BMO}. Indeed one interpolates between a non-BPS brane when $c_1=0$, and a BPS-brane when
$c_1 = \sqrt{5/2}$ ($k=0$).  The decay of the non-BPS brane can be
viewed here as the trajectory in the parameter space when $c_1\in
[0, \sqrt{5/2}]$. In the following the value of $c_1$ is not
specified, and we study the dynamics of a probe in the non-BPS
background for any $c_1$ (thus corresponding to any stage of the
decay).

We begin with the non-compactified case.

\section{Brane dynamics and mirage cosmology}

In this section we determine the dynamics of a probe BPS D3-brane
in the non-compact background spacetime (\ref{metricE}),
neglecting any back-reaction effects of the probe onto the
background spacetime.

We consider an infinitely straight brane, lying parallel to the
$x^{\mu}$ ($\mu=0,1,2,3$) axes, but free to move along the $r$ and
$\varphi^{q}$ ($q=5,\ldots,9$) axes. Its position, $X^{A}(t)$, is
therefore given by
\begin{equation}X^{\mu}=x^{\mu}, \qquad X^r(t)=R(t), \qquad X^{q}(t) =
\varphi^{q}(t). \label{static}
\end{equation}
%which is just the static gauge.
Below the Born-Infeld action is used to determine $R(t)$,
$\varphi^I(t)$, and we will see that unless the BPS probe brane
has sufficient angular momentum, it is attracted to the non-BPS
brane and so the naked singularity at $r=r_0$ is not shielded.

For an observer living on the brane, the brane motion
(\ref{static}) leads to a FRW universe with scale factor $S(\tau)$
where $\tau$ is brane time.  This effective `mirage' cosmology is
due only to brane motion and not brane self-gravity \cite{KK}.  In
the resulting  Friedmann equation, which we determine, the brane
motion leads to `dark fluid' terms whose equation of state,
$\omega$, is calculated \cite{SP}. We also show that $0 \leq
S(\tau) \leq 1$ when $c_2 \geq 1$ so that inflation may not occur
on the brane in this case.

A crucial component which is missing in this mirage cosmology
approach is brane self-gravity.  This is included in section
\ref{phil} where we compactify the background spacetime, and in
that case we will see that inflation may occur.  The scale factor
$S(\tau)$ mentioned above and discussed in this section will still
play a r\^ole when self-gravity is present.

\subsection{Action and dynamics}

We determine the probe BPS brane dynamics by solving the
equations of motion coming from the action 
\begin{equation}
S = S_{DBI} + S_{WZ} = -T  \int d^{4} x e^{-\phi} \sqrt { - \det
(\gamma_{\mu \nu} + (2 \pi \alpha')F_{\mu \nu} - \hat{B}_{\mu \nu}
) } - q T \int \hat{C}_4 \label{action}
\end{equation}
where $T$ is the brane tension, $\alpha'$ the string tension, and
$q=(-)1$ for a BPS (anti-)brane. In the DBI term, $\gamma_{\mu
\nu}$ is the induced brane metric
\begin{equation}
\gamma_{\mu \nu} =
\partial_\mu X^{A} \partial_\nu X^{B}  \bar{G}^s_{A B}(X)
\end{equation}
where the superscript $s$ refers to the string frame.  For the
remainder of this section we ignore the pull-back of the
Neveu-Schwarz anti-symmetric two-form $\hat{B}_{\mu \nu}$ as well
as the worldvolume anti-symmetric gauge fields $F_{\mu \nu}$.
These will be considered in section \ref{phil} when we discuss
bi-metric theories. The WZ term is
\begin{equation}
\int \hat{C}_4 = \frac{1}{4!} \int d^4 x \epsilon^{\mu \nu \rho
\sigma}
\partial_{\mu}X^{A} \partial_{\nu}X^{B} \partial_{\rho}X^{C}
\partial_{\sigma}X^{D} C_{ABCD}
\end{equation}
with $\epsilon^{0123}=1$.

In the static gauge (\ref{static}), the induced metric is
\begin{equation}\gamma_{00} = e^{\phi/2} \left(\bar{G}_{00} +
\bar{G}_{rr}\dot{R}^2 +
\bar{G}_s \dot{\varphi}^2 \right)
\qquad, \qquad \gamma_{ij} = e^{\phi/2}  \bar{G}_{ij}
\end{equation}
where $\cdot = d/dt$ and $\dot{\varphi}^2 =
h_{pq}\dot{\varphi}^{p} \dot{\varphi}^{q}$ with $h_{pq}$ the
metric on the 5-sphere. Since the functions appearing in action
(\ref{action}) depend on $R,~\varphi^{q}$, and hence only on $t$
(by virtue of (\ref{static})), one can therefore define a
Lagrangian ${\cal L}$ through
\begin{equation}
S = T V_3 \int dt {\cal L}
\end{equation}
where $V_3 = \int d^3 x$ is the (infinite)
spatial volume of the probe, and
\begin{equation}{\cal L} = - \sqrt{a+ b\dot{R}^2 + c \dot{\varphi}^2 } + e
\end{equation}where
\begin{equation}a = e^{8A} , \qquad b = - e^{2(3A + B)} = \frac{c}{R^2} , \qquad e
= - q e^{\Lambda}. \label{abcddef}
\end{equation}

This Lagrangian defines a conserved positive energy $E$ and
angular momentum $l$ of the brane around the $S^5$ which are
given by \cite{KK}
\begin{eqnarray}
E &=& \frac{\partial {\cal L}}{\partial \dot{R}} \dot{R} +
\frac{\partial {\cal L}}{\partial \dot{\varphi}^{p}}
\dot{\varphi}^{p} - {\cal L} \nonumber
\\
l^2 & = & h_{pq}\frac{\partial {\cal L}}{\partial
\dot{\varphi}_p} \frac{\partial {\cal L}}{\partial
\dot{\varphi}_q}. \nonumber
\end{eqnarray}
Hence
\begin{eqnarray}
\dot{\varphi}^2 & = &
 \frac{a^2 l^2}{c^2(E+e)^2}
\nonumber
\\
\dot{R}^2 & = & -\frac{a}{b} \left[ 1 + \frac{a}{c} \frac{ (l^2
- c)}{(E+e)^2} \right]. \label{Rdot}
\end{eqnarray}

Since $a,b,c,$ and $e$ are known functions of $R$, these equations can
now be solved to determine the brane dynamics as seen by an observer in
the bulk with time coordinate $t$.
For an observer living
on the brane, however, the line element is
\begin{eqnarray}
ds_{3}^2 &=& \gamma_{\mu \nu}dx^{\mu}dx^{\nu} \nonumber
\\
& = & \gamma_{00}(t) dt^2 +  \gamma_{ij} dx^{i} dx^{j} \nonumber
\\
& \equiv & - d\tau^2 + S^2(\tau) d{\bf x}^2 \label{Sdef}
\end{eqnarray}
where the brane time $\tau$ and the scale factor $S(\tau)$ are given by
\begin{equation}d\tau^2 = - \gamma_{00}(t) dt^2 \qquad , \qquad S^2 = e^{\phi/2} e^{2A}
\label{eta}
\end{equation}Thus the observer on the brane appears to be living in a FRW
universe with scale factor $S(R(\tau))$ arising from the brane
motion.  Notice that there is no term in action (\ref{action}) of
the form $\int d^4 x \, \sqrt{-\gamma}{\cal R}_{\gamma}$ (where
${\cal R}_{\gamma}$ is the 4-dimensional Ricci scalar). Thus the
above scale factor is only due to brane motion and not due to
self-gravity on the brane: in section 4 we show how self-gravity
can be included.  In that case the scale factor $S(\tau)$ will
still be important.

From Eqs.~(\ref{eta}) and (\ref{Rdot}) we find
\begin{equation}
d\tau^2 =
%\frac{e^{\phi/2}}{g_d^p}(a +
%b \dot{r}^2 + c h_{ab}\dot{\phi}^{a} \dot{\phi}^{b}) dt^2 =
 e^{-6A + \phi/2} \frac{a^2}{(E+e)^2} dt^2
\label{teta}
\end{equation}
so that $R^{'2}$ (where $'=d/d\tau$) is given by
\begin{equation}
R^{'2} = - \frac{1}{abc}e^{6A -\phi/2} \left[ c(E+e)^2 + a (l^2 -
c) \right].
\label{Rdash}
\end{equation}
Finally the Friedman equation on the brane is
\begin{equation}
H^2 = \left( \frac{1}{S} \frac{dS}{d\tau} \right)^2 = R^{'2} \left[
\frac{dA}{dR} + \frac{1}{4} \frac{d\phi}{dR} \right]^2 \equiv
\frac{8 \pi}{3} \rho_{{\rm eff}} \label{Hdef}
\end{equation}
which defines an effective energy density.  The corresponding
equation of state, which we determine in section 3.3, is
\begin{equation}
\omega = \frac{p_{{\rm eff}}}{\rho_{{\rm eff}}} = - \left( 1 +
\frac{1}{3} \frac{S}{H^2} \frac{\partial H^2}{\partial S} \right).
\label{omega}
\end{equation}
Clearly in order to find $\omega$, we need first to determine the
brane dynamics, $R(t)$.  This will depend on the brane energy $E$
and angular momentum $l$ as well as the parameters $c_1$, $c_2$
characterising the non-BPS bulk.

\subsection{Brane dynamics}

%In order to study the brane dynamics, i
It is helpful to use the analogy with particle dynamics and to
study the brane motion via the two effective potentials
\begin{equation}
V^{t}(R,l,E) = E - \frac{1}{2}\dot{R}^2 , \qquad
V^{\tau}(R,l,E) = E - \frac{1}{2}{R'}^2 \label{Vdef}
\end{equation}
where $\dot{R}$ and $R'$ are given respectively in equations
(\ref{Rdot}) and (\ref{Rdash}).  The first potential, $V^t$,
describes the brane motion as seen by a bulk observer.  The
second, $V^{\tau}$, determines the brane dynamics as seen by an
observer on the brane. Allowed regions of $R$ are ones for which
%$\dot{R}^2, ~{R'}^2 > 0$, so that
$V^{t,\tau}(R,l,E) \leq E$.  Notice the dependence of the
potentials on $E$ and $l$.

When the probe is very far from the non-BPS brane, $R \rightarrow
\infty$, it is straightforward to show from (\ref{Rdot}) and
(\ref{Rdash}) that for all $c_1$, $c_2$ and $l$,
\begin{equation}
\dot{R}^2 \rightarrow 1-\frac{1}{E^2}, \qquad {\rm and} \qquad
R^{'2} \rightarrow E^2 - 1.
\end{equation}
Hence both in background time $t$ and brane-time $\tau$,  the
probe will be unable to escape to infinity if $E<1$.  In other
words, for $E<1$ the BPS probe brane is always bound to the
non-BPS brane and, depending on $l$ (see below) will eventually
be absorbed by it. On the other hand, if $E>1$ the probe will be
able to escape. The limiting case is when $E=1$, for which the
kinetic energy of the probe vanishes at infinity where there is no
force on the probe. In the BPS limit (of the non-BPS background,
as discussed in \cite{BMO}), both $\dot{R}$ and $R'$ vanish for
all $R$ when $E=1$, so that $R$ is a flat direction, as expected.

To determine the brane dynamics for $r_0 \leq R < \infty$ consider
first $V^{\tau}$ which is the relevant potential for an observer
living on the brane.  Very similar comments to the ones below also
hold for $V^{t}$.  We focus on the neutral limit, $c_2=1$, which
will be of most interest when self-gravity is included in section
\ref{phil}.  (Notice that when $c_2=\pm 1$ then $e=0$ and the
coupling between the BPS probe and the bulk RR field vanishes.
Below we see that for $c_2=-1$ the scale factor $S(\tau)$ is badly
defined.)  Using (\ref{Vdef}) and (\ref{Rdash}) as well as the
definitions of $a$, $b$ and $c$ in (\ref{abcddef}) gives
\begin{equation}
V^{\tau}(R,l,E) = E - \frac{1}{2} \frac{
(f_+)^{(3k+c_1-1)/2}}{(f_-)^{(3k+c_1+1)/2}} \left[ E^2 - \left(
\frac{f_-}{f_+} \right)^{2k}  - \frac{l^2}{r^2 } \frac{
(f_-)^{k-1/2}}{(f_+)^{k+1/2}} \right]. \label{Vtau}
\end{equation}

When the probe has zero angular momentum, $l=0$, and when
$k>0$, equation (\ref{Vtau}) gives $V^{\tau}(r_0,0,E) \rightarrow
- \infty$. Thus for all $E$ there is an infinitely deep potential
well at $R=r_0$: the probe is attracted to the singularity which
is therefore not protected. If $E<1$ then the probe is bound in
the region
$$
r_0 \leq R \leq \bar{r} = r_0 \left( \frac{1 + E^{1\over {k} }}{1
- E^{1 \over {k}}} \right)^{1/4}.
$$

When $l \neq 0$, a centrifugal potential can develop and this
may shield the singularity.  From (\ref{Vtau}) notice that if $k <
1/2$ then $V^{\tau}(r_0,l,E) \rightarrow + \infty$ for all $E$
so that the singularity is shielded.  If $k \geq 1/2$, however, $V
^{\tau}(r_0,l,E) \rightarrow - \infty$ so that the probe may
reach the singularity. Despite that some regions of $R$ may be
excluded even for $k \geq 1/2$, i.e. if $l$ is sufficiently
large $l
> l_c$ where $l_c$ is a critical angular momentum which
depends on $E$, $r_0$ and $c_1$,  then one can show that there are
indeed such excluded regions \cite{SP}.  These results are shown
in figure 1.
\FIGURE{
$$\epsfbox{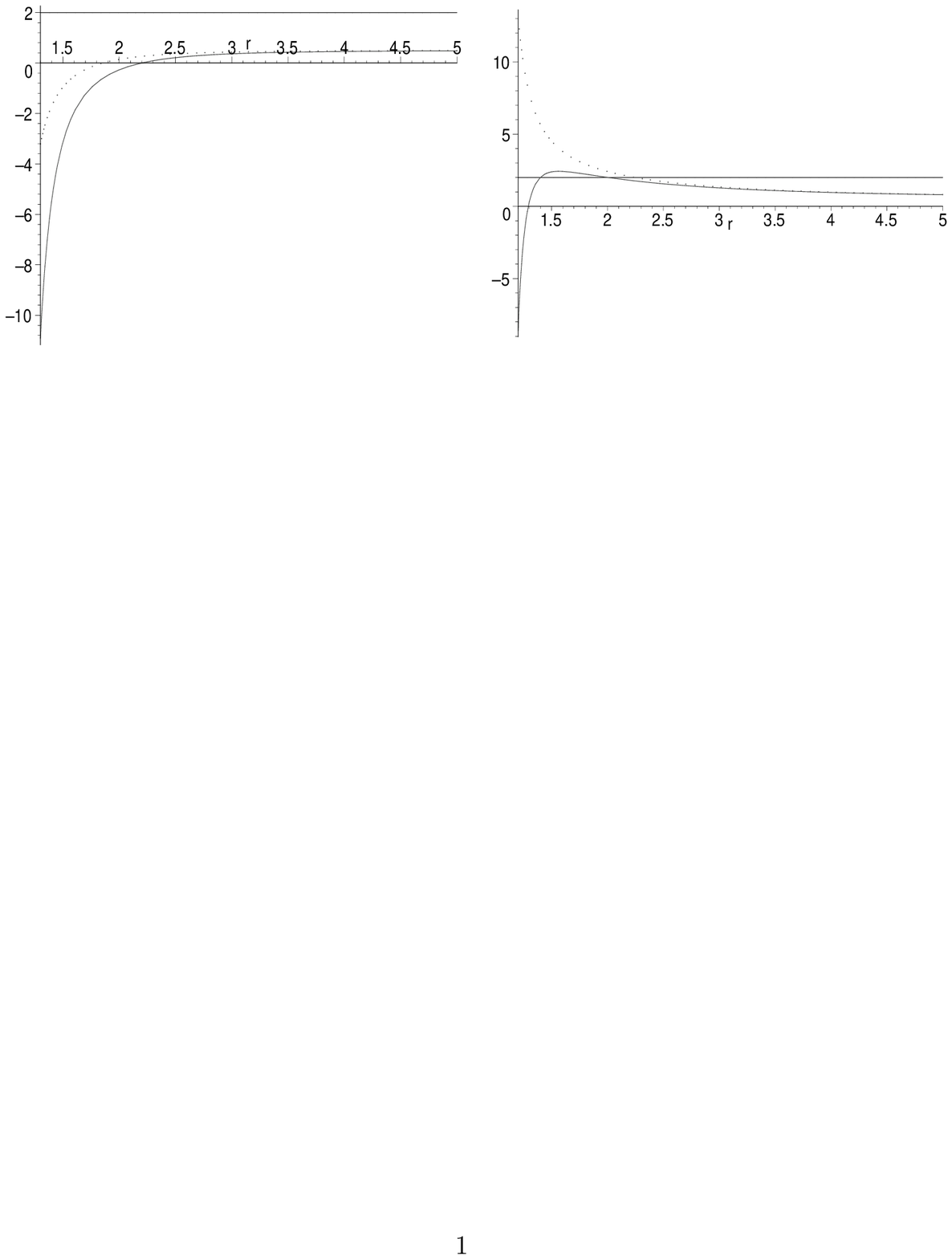}$$
\caption{Graphs of
$V^\eta(r,l,E)$ as a function of $r$.
  For simplicity we have chosen the parameters
 to be $q=1=\tilde{\eta}$ and $r_0=c_2=1$.  The probe energy is $E=2$
 corresponding to the horizontal line.  The LH figure is for a
 probe with zero angular momentum $l=0$ --- the dotted curve
 has $k=0.44 < 1/2$,
% $c_1=1.52 > 3/2$,
and the other $k=3/2$.
% $c_1 = 0.5 < 3/2$.
In the RH figure, $l= 4 > l_c $.}}

\subsection{Mirage cosmology}

How is this dynamics reflected in the Friedmann equation on the
brane?  From Eqn.~(\ref{eta}), the brane scale factor $S(\tau)$ is
given by
\begin{equation}S(\tau)^4 = e^{\phi} e^{4A} = \left( \frac{f_-}{f_+} \right)^{c_1}
\frac{1}{Y},
\end{equation}where $Y$ is a function of $c_2$.  If $c_2 < -1$, there are
regions of $R$ for which $S^4 < 0$ thus leading to an imaginary
scale factor: these regimes cannot be physical and so we take $c_2
\geq 1$. In that case the scale factor is bounded by
$$
0
\leq S \leq 1
$$
since $S=0$ as $R \rightarrow r_0$ and $S=1$ as $R \rightarrow
\infty.$  Thus there can be no inflation on the brane in this
mirage cosmology approach.

Again we consider the neutral limit $c_2=+1$.  Then
\begin{equation}S^4 = x^{c_1 + k}  \qquad {\rm where} \qquad x = \frac{f_-}{f^+}
\label{xS}.
\end{equation}
Furthermore, by definition of $H$ in (\ref{Hdef}) one has
\begin{equation}
H^2 = \frac{1}{2} \frac{(c_1 + k)^2}{r_0^2}
\frac{(1-x^2)^{5/2}}{x^{\frac{5+3k+c_1}{2}}} \left[ E^2 - x^{2k} -
\frac{l^2}{r_0^2} \frac{(1-x^2)^{1/2}}{2x^{1/2}} \right]
\label{Hres}
\end{equation}which, combined with (\ref{xS}) gives $H^2(S)$.  Hence $\omega$
defined in (\ref{omega}) can be calculated.  There are two
difficulties though:  firstly $H^2(S)$ is not a sum of terms of
the form $S^{-n}$ for some integer $n$, since functions such as
$(1-x^2)^{5/2} = (1-S^{8/(k+c_1)})^{5/2}$ appear in $H^2$.
Secondly, $c_1$ can take any continuous value between $0$ and
$\sqrt{5/2}$ and so $\omega$ can also take any value.

To overcome the first problem consider the limit $S \ll 1$ (and
hence $R = r_0 + \epsilon$).
%((((....actually we'd be more interested in the opposite regime))))).
A Taylor expansion of
$H^2$ then contains terms of the form
\begin{equation} x^{(\alpha k + \beta c_1 + \gamma)/2} =
S^{\frac{2 (\alpha k + \beta c_1 + \gamma)}{c_1+k}}
 \label{ss}
\end{equation}for integer $\alpha,~\beta$ and $\gamma$. Surprisingly two values
of $c_1$ are picked out for which this expansion of $H^2$
contains terms in $S$ to an integer power, they are\footnote{These
two values do not seem to correspond to any special case of the
non-BPS background metric (\ref{metricE}).}
\begin{equation}c_1 - k = \pm 1 \qquad \Longrightarrow \qquad c_1 = 3/2, \; \;
1/2. \label{c1spec}
\end{equation}
When $c_1 = 3/2$, the Friedmann equation contains `dark fluids'
with equation of state $\omega = 2,2/3,-2/3$.   These are sourced
by the angular momentum $l$ of the brane.  There are also
terms with $\omega = 5/3,-1$ which are sourced by the energy $E$
of the brane, and terms in $\omega = 1,~1/3$ which are independent
of $E$ and $l$.  Note that there is a term behaving as
radiation, as well as a cosmological constant term, but there is
none behaving as matter.

For $c_1=1/2$ the main difference is that there is now a dark
`matter' term $\propto S^{-3}$ and this is sourced by the angular
momentum $l$.  As far as we are aware, this is the first
background for which such a dark `matter' term appears in mirage
cosmology.

\section{Non-BPS Brane Inflation}
\label{phil}

In the previous section we described the brane dynamics in
non-compact space-time.  Now we concentrate on the low energy
dynamics of the same system, i.e.\ a probe brane in the vicinity
of a non-BPS brane, when the six extra dimensions have been
compactified. We do not assume any particular form for the
compactification manifold --- on the contrary we only need to
consider a patch where the non-BPS brane and the probe are
embedded. In that patch the geometry is curved by the non-BPS
brane while the effective field theory on the probe brane is
subject to two competing effects. Firstly gravity now propagates
on the brane, hence modifying the induced metric from the
previously studied mirage cosmology case. Secondly the motion of
the probe in the background of the non-BPS brane modifies the
geometry of the probe. The combined effect can be understood  by
analysing the equations of motion of the low energy effective field
theory below the compactification scale.

\subsection{Gravitons on the Brane}

Consider now the 10d metric (in the Einstein frame)
\begin{equation}
ds_E^2=e^{2A}g_{\mu\nu}dx^{\mu}dx^{\nu}+e^{2B}(dr^2 + r^2
d\Omega_5^2) \equiv G_{AB} dx^{A} dx^{B} \label{metric10}
\end{equation}
where $g_{\mu\nu}$ is any 4d metric, as opposed to the Minkowski
metric considered in (\ref{metricE}).
For simplicity we now focus on the neutral case,
$|c_2|=1$, in which case $F^2=0$ from equation (\ref{lambda}).
We will determine the equations of motion for $g_{\mu\nu}$
and show that one can construct a whole family of solutions.
The previously described solution $g_{\mu\nu}=\eta_{\mu\nu}$ is only
the simplest case. As long as we do not compactify six dimensions, the
dynamics of $g_{\mu\nu}$ are frozen due to the infinite effective four
dimensional Planck mass, i.e. four dimensional gravity decouples.
In the previous section we have considered the mirage dynamics where
the fluctuations of the four dimensional metric $g_{\mu\nu}$ have been
frozen. As soon as we compactify, the effective four dimensional
Planck mass becomes finite allowing one to study the gravitational
dynamics of $g_{\mu\nu}$.

 Using the relation
\begin{equation}
^{10}{\cal R}(G) = \; ^4 {\cal R}(g) e^{-2A} + \; ^{10}{\cal
R}(\bar{G}), \label{Ricci}
\end{equation}
it follows that the 10d SUGRA action for the metric
(\ref{metric10}) reduces to
\begin{equation}
S_{\hbox{sugra}}= \frac{1}{2\kappa_{10}^2}\int d^{10}x
\sqrt{-\bar G}\sqrt{-g}e^{-2A} \ ^4{\cal R}(g)
\end{equation}
(see also section 4.2). From this action, the equations of motion
for $g_{\mu\nu}$ reduce to the 4d Einstein equations
\begin{equation}
^4{\cal R}_{\mu\nu}-\frac{^4{\cal R} }{2}  g_{\mu\nu}=0.
\end{equation}
Hence
the metric $g_{\mu\nu}$ can be any four dimensional solution of
the vacuum Einstein equations --- a Schwarzchild metric, for
instance, leads to a solution of the 10d equations of motion.

Now consider small perturbations of the flat metric
$\eta_{\mu\nu}$, i.e. $g_{\mu\nu}=\eta_{\mu\nu}+ h_{\mu\nu}$. This
describes the moduli space of solutions where $h_{\mu\nu}$
satisfies $\eta^{\mu\nu}h_{\mu\nu}=0,\
\partial_{\mu}h^{\mu}_{\nu}=0$ and is massless
\begin{equation}
{}^4 \Box h_{\mu\nu}=0.
\end{equation}
Hence massless gravitons $h_{\mu\nu}$ can propagate on the brane.
Nevertheless, in this uncompactified case the effective four
dimensional Planck mass, defined through
\begin{equation}
\frac{1}{2\kappa_4^2}  = \frac{1}{2\kappa_{10}^2} \int
\sqrt{-\bar{G}} e^{-2A} dr d\Omega_5 \label{Meff4}
\end{equation}
diverges, implying the decoupling of self-gravity on the brane. In
the remainder of this section we therefore consider the
compactified case in which the compactification is over a large 6d
manifold ${\cal M}$, and gravity does not decouple anymore.

\subsection{Effective 4D action with self-gravity}

Here we compactify the theory and study the effective field theory at low
energy below the compactification scale
--- as explained above, this should contain
self-gravity. The low energy action therefore comprises two terms:
first there is the action obtained by dimensional reduction of the
warped background created by the non-BPS brane. Second we include
the Born-Infeld action (\ref{action}) reflecting the low energy
dynamics of the transverse modes to the probe.
In order to maintain the neutrality of the background we will focus on
the
case
$\vert c_2\vert =1$.

Before doing so, let us discuss the compactification in general.
Our starting point is the metric (\ref{metric10}) where $g_{\mu \nu
}dx^{\mu}dx^{\nu} = \eta_{ab} d\xi^{a} d\xi^{b}$ in a local
frame $d\xi^a$.
We make
the ansatz that both the form $F$ as well as the dilaton $\phi$
coincide with the solution obtained previously in the local frame
specified by $d\xi^a$.  This amounts to specifying that locally the
solution of the previous section is valid. Hence
\begin{equation}
F_{r \mu_0\dots \mu_{3}}dr\wedge dx^{\mu_0}\wedge \dots\wedge
dx^{\mu_{3}}= \Lambda' e^{\Lambda}dr\wedge d\xi^{a_0}\wedge
\dots\wedge d\xi^{a_{3}}
\end{equation}
It is through $g_{\mu \nu}$ that we will be able to describe
self-gravity on the brane.
The dynamics at low energy are determined by
\begin{eqnarray}
S &=&
\frac{1}{2 \kappa_{10}^2} \int d^{10}x \; \sqrt{-{G}} \left(
^{10}{\cal R}(G) - \frac{1}{2} G^{\mu \nu}
\partial_{\mu} \phi
\partial_{\nu} \phi  - \frac{1}{2 \cdot 5!} F^2
\right) \nonumber
\\
&-& T \int d^4 x e^{-\phi} \sqrt{-\gamma} - T q \int d^4 x C_4
\label{stuff}
\end{eqnarray}
where the first term is the bulk 10d supergravity action, and the
second is the DBI action (\ref{action}).  We will integrate out
the above action over the 6 extra dimensions which are
compactified on ${\cal M}$, and hence obtain an effective 4D
action which includes brane self-gravity through the appearance of
the 4D Ricci scalar $^{4}{\cal R}(g)$. We assume that the manifold
${\cal M}$ is fixed, i.e. the moduli have been stabilized by an
unknown
mechanism.   On using
(\ref{Meff4}), the bulk part of action (\ref{stuff}) becomes
\begin{equation}
S_{\rm bulk}
 =\int d^4 x \sqrt{-g} \left[ \frac{1}{2 \kappa_{4}^2} \;
^{4}{\cal R}(g) \right] - \frac{1}{4 . 5! \kappa_{10}^2}\int d^4 x
\sqrt{-g} \left( \int_{\cal M} dr d\Omega_5 \sqrt{-\bar{G}}
 F^2\right )
 \label{bulkaction}
\end{equation}
where the 4D planck mass is defined analogously to (\ref{Meff4}),
namely
\begin{equation}
\frac{1}{2\kappa_4^2}  = \frac{1}{2\kappa_{10}^2} V_6
\end{equation}
where $V_6$ is the now finite volume of the 6d compact space;
\begin{equation}
V_6\equiv R_c^6=\int_{\cal M} \sqrt{-\bar{G}} e^{-2A} dr
d\Omega_5.
\end{equation}
Notice that $\kappa_4^2$ is a function of $c_1$, and also that
\begin{equation}
F^2  =  - 5! \left( \Lambda' e^\Lambda \right)^2 e^{-8A-2B} =  -
5! \frac{ 64 \xi^2 k^2 (c_2^2-1) r_0^8}{Y^{5/2} r^{10} f_+^{5/2}
f_-^{5/2}}.
\end{equation}
Notice that the factors of $A$ and $B$ appear due to the non-trivial contraction with the background metric.
This defines the gravitational part of the effective action. Notice
that there is a negative cosmological constant related to the
background field strength.
Hence in the neutral case ($c_2 = \pm 1$), both $F^2$ and the
coupling of the probe to the four form vanish leading to the absence
of the cosmological constant. On the other hand,
in the charged case the resulting action contains a negative
cosmological constant leading to AdS$_4$ in the absence of the
probe.

Now consider the brane part of the action. We work in the isotropic gauge
where
\begin{equation}
X^{\mu} = x^{\mu}
\end{equation}
but as opposed to section 2 the brane is now allowed to be bent.
Hence
\begin{equation}
X^{r}=R(x^\mu), \ X^{p} = \varphi^{p}(x^\mu) ,
\end{equation}
and the induced metric is thus
\begin{equation}
\gamma_{\mu \nu}=  G^s_{\mu \nu} + G^s_{rr} \frac{\partial
R}{\partial x^{\mu}} \frac{\partial R}{\partial x^{\mu}} +
G^s_{pq} \frac{\partial \varphi^{p}}{\partial x^{\mu}}
\frac{\partial \varphi^{q}}{\partial x^{\nu}}. \label{bi}
\end{equation}
 We will focus on the motion in
the $r$ direction (and so set $l = 0$ in the notation of the
previous section). In the DBI action, we need to consider the
expansion of $\sqrt{-\gamma}$: if we assume that the brane moves
slowly  so that the square root part of the DBI action can be
expanded one gets, to leading order,
\begin{equation}
\sqrt{-\gamma}=e^{4A} e^{\phi}  \sqrt{-g} \left[ 1 + \frac{1}{2}
e^{2B-2A}
  g^{ab} \frac{\partial R}{\partial x^{a}} \frac{\partial R}{\partial
x^{b}} \right].
\end{equation}
(Below we will see that $R$ is very closely related to the
inflaton on the brane, and that this assumption of a slowly moving
brane is consistent with the slow-roll conditions.) Thus
\begin{equation} S_{\rm brane} = - T \int d^4 x \sqrt{-g}
e^{4A} \left[ 1 + \frac{1}{2} e^{2B-2A} g^{ab} \partial_a R
\partial_b R \right] - q T \int d^4 x \sqrt{-g} e^{\Lambda}.
\label{braneaction}
\end{equation}
This is a non-linear sigma model for $R$ with a non-vanishing
potential
\begin{equation}V(R) = T \left( e^{4A} - q e^{\Lambda} \right)
\end{equation}due to the absence of supersymmetry.\footnote{Note that this
potential is not directly related to the effective potentials
$V^{\tau,t}$ discussed in the previous section.} This potential is valid generically for any
motion of a $D3$ probe in a non-trivial gravitational background. In the
supersymmetric case the contributions from the Wess-Zumino
coupling and the DBI cancel. Here the absence of cancellation
leads to a non-trivial dynamics for the interbrane distance.
Notice also the wave-function normalization term
\begin{equation}Z(R) = T e^{2B+2A}.
\end{equation}

It is convenient to use fields with mass dimension one by putting
\begin{equation}
\phi = \frac{R}{l_s^2}
\end{equation}
where $l_s$ is the string length scale.  We now focus on the
neutral case so that it follows from (\ref{bulkaction}) and
(\ref{braneaction}) that the effective 4D action is given by
\begin{equation}
S^4_{\rm eff} = \int d^4 x \sqrt{-g} \left(
 \frac{1}{2 \kappa_{4}^2} \;
^4{\cal R}(g)  - \frac{1}{2} Z(\phi) \partial_\mu \phi
\partial^{\mu} \phi - V(\phi) \right).
\label{Seff}
\end{equation}
Below we study this action  and  we will be interested in the
supergravity regime where the two branes are far apart $R\gg r_0$
leading to
\begin{eqnarray}
V(\phi) &=& T \left( 1 - \frac{2k r_0^4}{l_s^8 \phi^4} \right)
\nonumber
\\
Z(\phi) &=& l_s^4 T  \nonumber
\end{eqnarray}
where we have neglected higher order terms than $1/\phi^4$  in $Z$.

In the next section we will analyse cosmological solutions to the
equations of motion coming from action (\ref{Seff}).  Before that,
notice an intriguing point  --- the induced metric on the
brane $\gamma_{\mu \nu}$ differs from the metric $g_{\mu\nu}$
which is responsible for Einstein gravity at low energy.
Furthermore matter couples to the induced metric as can be seen by
including a matter term in the DBI corresponding to the pull-back
of the $U(1)$ field strength
\begin{equation}
F_{\mu \nu}= F_{AB}\frac{\partial X^{A}}{\partial x^\mu}
\frac{\partial X^B}{\partial x^\nu}.
\end{equation}
Action (\ref{action}) leads to the kinetic terms
\begin{equation}
(2 \pi \alpha')^2 \frac{T}{8}\int d^4x \sqrt{-\gamma}F^{\mu \nu
}F^{\rho \delta}\gamma_{\mu \rho}\gamma_{\nu \delta}
\end{equation}
so that matter does indeed couple to the induced metric in the
string frame. Thus we have a bi-metric theory \cite{YI,YII}
where
\begin{equation}
\gamma_{\mu \nu} = e^{\phi/2} e^{2A} g_{ab} + e^{\phi/2} e^{2B}
\partial_a R \partial_b R.
\label{gammag}
\end{equation}
As soon as the brane stops moving, the theory stops being a
bi-metric theory.

\subsection{Inflationary solution and induced metric}

We search for inflationary solutions of (\ref{Seff}) of the form
\begin{equation}
g_{ab} = {\rm diag}(-1,a^2(t),a^2(t),a^2(t)).
\end{equation}
It should be noted, however, that $a$ is not the scale factor as
seen by an observer living on the brane.  The metric on the brane
is just the induced metric $\gamma_{\mu \nu}$ as discussed above
and as given in (\ref{gammag}). Assuming now, for simplicity, that
$R=R(t)$ only, the metric on the brane is therefore given by
\begin{eqnarray}
ds^2 &=& dt^2 (e^{\phi/2} e^{2A} g_{00} + e^{\phi/2}
e^{2B}\dot{R}^2) + e^{\phi/2} e^{2A}g_{ij}dx^{i} dx^{j} \nonumber
\\
& \equiv& -d\tau^2 + \tilde{a}^2(\tau) d{x}^2
\nonumber
\end{eqnarray}
where $\cdot = d/dt$ and
\begin{equation}
\tilde{a}(\tau) \equiv S(\tau) a(\tau).
\end{equation}
It combines the mirage cosmology brane factor $S^2(\tau) = e^
{\phi/2}e^{2A}$ which is due to the motion of the brane, and $a$
which arises from the self-gravity on the brane. As usual,
inflation will lead to an exponential expansion on the brane $a
\sim e^{Ht}$.

Since matter on the brane couples to $\gamma_{\mu \nu}$, whilst
gravity couples to $g_{ab}$, predictions regarding the CMB and the
spectrum of perturbations should in principle be made by
calculating with respect to the scale factor $\tilde{a}$ and
brane-time $\tau$, since particle and photon geodesics are
governed by the metric $\gamma_{\mu \nu}$. Thus one should replace
$g_{ab}$ in the effective action (\ref{Seff}) by $\gamma_{\mu
\nu}$, i.e.\ inverting (\ref{gammag}). This, however, would lead
to a rather complicated scalar-tensor action, and in that frame
the equations of motion are not transparent. Moreover the
generation of quantum fluctuations is easier to understand in the
Einstein frame where all the usual formalism applies. We will
comment on the coupling to matter later and from now on work in
the Einstein frame with action (\ref{Seff}) and time $t$ (as
opposed to brane time $\tau$). Eventually, in the inflationary
phase, the difference between $t$ and $\tau$ can be seen to be
negligible due to the slow roll condition. Similar actions were
taken as the starting point of references
\cite{DI,DII,BUI,BUII,D,SH}.

The equations of motion coming from the effective action are
\begin{eqnarray}
\ddot{\phi} + 3 H \dot{\phi} + \frac{1}{2} \left( \frac{Z'}{Z}
\right) \dot{\phi}^2 + \frac{V'}{Z} &=& 0, \nonumber
\\
H^2 = \left( \frac{\dot{a}}{a} \right)^2 = \frac{\kappa_4^2}{3}
\left( \frac{1}{2} Z \dot{\phi}^2 + V \right)&\ & \nonumber.
\end{eqnarray}
Assuming that
\begin{equation}
\frac{1}{2} Z \dot{\phi}^2 \ll V,\ \qquad \ddot{\phi} \ll 3 H
\dot{\phi}
\end{equation}
i.e.\ in the slow roll regime, yields
\begin{equation}
H^2 = \frac{\kappa_4^2}{3} V \simeq \frac{\kappa_4^2 T}{3}
\end{equation}
leading to
\begin{equation}
a(t) = a(t_0)e^{\sqrt{\frac{\kappa_4^2 T}{3}}(t-t_0)}.
\end{equation}
As
\begin{equation}
\dot{\phi} = - \frac{V'}{3HZ} \label{phidot}
\end{equation}
where $V'$ is positive, it follows that during inflation $\phi$
(or equivalently $R$) decreases.  Notice that the second scale
factor $S$ actually decreases as $R$ decreases.  Since the scale
factor on the brane is actually $Sa$, the overall effect will
still be one of inflation since $S$ is only a powerlaw in $t$.
From (\ref{phidot}) we find that
\begin{equation}
\phi^6 = \phi_0^6 - Q (t-t_0)
\end{equation}
where
\begin{equation}
Q = \frac{48 k r_0^4 }{ \sqrt{3} l_s^{12}  \kappa_4 \sqrt{T}}.
\end{equation}

The slow-roll regime is valid provided that
\begin{equation}
\eta = M_{pl,4}^2 \frac{V''}{V} \ll 1 ,\ \qquad \epsilon =
\frac{1}{2} M_{pl,4}^2 ( \frac{V'}{V} )^2 \ll 1.
\end{equation}
Here $\epsilon$ is always smaller than $\eta$ so we only consider the
first
slow roll condition.
Imposing the first condition gives
\begin{equation}
|\eta| = M_{pl,4}^2  \frac{40 k r_0^4}{l_s^8 \phi^6} \ll 1
\end{equation}
or equivalently
\begin{equation}
k \ll\frac{1}{40} \frac{r_{ini}^6}{M_{pl,4}^2 l_s^4 r_0^4}
\end{equation}
where we have taken the initial brane position $R =r_{ini} \gg
r_0$ consistently with our supergravity approximation. Using now
\begin{eqnarray}
T &=&  \frac{1}{g_s (2 \pi)^3 l_s^4}
\nonumber
\\
\kappa_{10}^2 &=& \frac{1}{2} (2\pi)^7 l_s^8
\end{eqnarray}
gives
\begin{equation}
k\ll \frac{(2\pi)^8}{20}\frac{l_s^4}{r_0^4}\left(
\frac{r_{ini}}{R_c} \right)^6.
\end{equation}
Thus $k$ is undetermined as long as $r_0$ remains a free parameter
of the supergravity approximation.

In order to specify the range of validity of inflation in the
non-BPS scenario, we need to compare the time taken by the non-BPS
to decay with the typical inflationary time scale.  Inflation
typically takes place within
\begin{equation}
t_{infl}=\frac{1}{H}
\end{equation}
whereas the decay time for the non-BPS brane is assumed to be
infinite in the supergravity approximation. Indeed, in the
approach we use here, the tachyon parameter $c_1$ is a constant
parameter and not a time-dependent modulus \cite{BMO}. A more
appropriate estimate of the decay time in the inflationary regime
can be obtained from
\begin{equation}
t_{decay}=\frac{1}{M_{ADM}}
\end{equation}
where $M_{ADM}$ is the ADM mass of the non-BPS brane \cite{BMO}.
When the brane is flat the ADM mass is infinite reflecting the
fact that it has a finite mass per unit volume
\begin{equation}
\frac{M_{ADM}}{V_3}=\frac{5}{2^7\pi^4}k_{max}\frac{r_0^4}{l_s^8}
\end{equation}
where $k_{max}=\sqrt{5/2}$. Now when the brane inflates, the
volume of space-time is of order of $V_3=H^{-3}$ implying that
\begin{equation}
t_{decay}=\frac{2^7 \pi^4}{5k_{max}}\frac{l_s^8}{r_0^4}H^3.
\end{equation}
Imposing now that $t_{decay}\gg t_{inf}$ leads to
\begin{equation}
r_0^4\ll \frac{1}{40 . 6^2 .
\pi^4}\frac{1}{g_s^2}\frac{1}{M_{pl,4}^4}.
\end{equation}
Notice that $r_0$ is linked to the 4d Planck mass.
In the following we will take $r_0=1/M_{pl,4}$ which satisfies the above
inequality at small string coupling.
This leads to
\begin{equation}
\frac{k}{k_{max}}\ll \frac{9}{2} (2\pi)^{12} g_s^2 l_s^4
M_{Pl,4}^4(\frac{r_{ini}}{R_c})^6
\end{equation}
We will discuss the range of values for $k$ in the next section.

\subsection{Reheating}

In this subsection we describe the end of the inflationary era and
its connection with the cosmological eras. First of all inflation
on the brane stops when $\eta\sim 1$. We find that
\begin{equation}
Q(t_{end}-t_0)=\frac{r_{ini}^6}{l_s^{12}}-\frac{40M^2_{pl,4}r_0^4}{l_s^8}.
\end{equation}
After the end of inflation the brane starts rolling fast towards
the non-BPS brane. When the  non-BPS brane decays the probe has to
be far enough from the non-BPS brane in order to trust the
supergravity approximation:  this leads to
\begin{equation}
g_s^2\gg
\frac{2^{21/2}\pi^5}{15}\frac{k}{k_{max}}\frac{l_s^{12}}{r_{ini}^6R^{6}}.
\end{equation}
This condition is always satisfied for the range of parameters
compatible
with the COBE normalization.
When the non-BPS brane decays, it leads  to bulk radiation in the
Minkowski vacuum. Moreover we know that $k_{decay}=0$ implying
that the potential vanishes altogether
\begin{equation}
V_{decay}=V_0 + T\equiv 0,
\end{equation}
where $V_0$ is the contribution at low energy to the tachyon
condensation process. It goes beyond the supergravity
approximation and should encompass the opens string degrees of
freedom leading to the decay of the non-BPS brane. The vanishing
of $V_0+T$ is the famous Sen conjecture for non-BPS branes
\cite{SI}. Here it can be viewed as the vanishing of the
cosmological constant at low energy due to the
overall supersymmetry of the configuration. The only trace of the
decay may appear from the coupling of the bulk radiation to the
brane YM fields leading to a radiation dominated era.

 Once the non-BPS brane decays, the brane stops moving
and starts expanding only due to the presence of matter on the
brane.
The background metric being flat Minkowski space, the ten dimensional
metric becomes
\begin{equation}
ds_{10}^2= dx_Adx^A -dt^2 +a^2(t) dx^idx_i
\end{equation}
where
$A=1\dots 6$ and $i=1\dots 3$. This implies that the induced
metric and the Einstein metric become equivalent. Moreover the six
extra dimensions play no role in the four dimensional dynamics. We
can therefore apply the usual formalism to compute the effect of
the initial quantum fluctuations on the CMB results. Indeed once inflation stops, the cosmology
of the $D3$ brane is solely dictated by the presence of matter on a static brane. In that case we can follow
the evolution of the brane through the usual cosmological eras and use the CMB results in order to calibrate the
underlying physics of the brane inflation.

In particular we can use the COBE normalization
\cite{SM,BE,LE,NE,PR} specifying th magnitude of the density
perturbations on large scales in order to evaluate the
compactification scale. This  gives
\begin{equation}
M_{pl,4}^{-3} \frac{V^{3/2}}{V'} \sim M_{pl,4}^{-3} \frac{V_6}{k
l_s^9} \sim 10^{-5}.
\end{equation}
Now
\begin{equation}
l_s^9 \sim  \left( \frac{V_6}{M_{pl,4}^2} \right)^{9/8},
\end{equation}
and on writing $M_c^6 = V_6^{-1}$, one has that
\begin{equation}
M_c  \sim 10^{12} {\rm GeV}.
\end{equation}
Using the relation between the four dimensional Planck mass, the compactification volume and the string scale we deduce that
\begin{equation}
M_s \sim M_{pl,4}^{1/4} M_c^{3/4} \sim 10^{14} {\rm GeV},
\end{equation}
so that the compactification scale is two orders of magnitude
smaller than the string scale. Notice that $r_0\ll R_c$ as it should in order to consider that the non-BPS brane
distorts only locally and weakly the geometry of the compact manifold ${\cal M}$.

Let us now consider the observable consequences of the primordial
inflationary era due to the presence of a  non-BPS brane. Two
crucial tests have to be overcome. First of all the number of
e-folds must be large enough. The number of e-folds
\begin{equation}
N=\int_{t_{ini}}^{t_{end}} H dt
\end{equation}
is determined by the initial position of the probe. Indeed
\begin{equation}
N= \frac{5}{6}\left(\frac{r_{ini}}{r_{end}} \right)^6.
\end{equation}
Using
\begin{equation}
r_{end}^6=40kr_0^4l_s^4 M^2_{pl,4}
\end{equation}
this implies that
\begin{equation}
r_{ini}^6=48N kr_0^4 l_s^4 M^2_{pl,4}.
\end{equation}
Taking $r_0$ to be of the order of $M_{pl,4}^{-1}$, i.e.\
 saturating its
upper bound,
leads to
\begin{equation}
\frac{r_{ini}}{r_0}\sim \left(\frac{M_{pl,4}}{M_s} \right)^{2/3}
\end{equation}
Notice that the ratio of the Planck scale to the string scale is
related to the COBE normalization, i.e.\ $10^{-5}$, we find that
\begin{equation}
\frac{r_{ini}}{r_0}\sim 10^3.
\end{equation}
Of course this justifies the supergravity approximation.
Notice that all the different scales of the problem have been determined from two inputs.
First of all the non-BPS brane scale $r_0$ is taken to be of the order the Planck scale, this to guarantee
that inflation stops well before the decay of the non-BPS brane. Then the COBE normalization fixes the volume
of the compactification manifold.

Now we can check that inflation takes place irrespective of the value $k$.
Hence we find that inflation is generic in the non-BPS scenario.

Within the supergravity approximation the spectral index of
fluctuations can be computed. It is given by \cite{SH}
\begin{equation}
n_s-1= -\frac{5}{3N}.
\end{equation}
Of course this is not a prediction of the model as $N$ is left undetermined,
i.e. $N$ depends on the initial position of the probe.
The model does not explain why the initial conditions should be such that
$N=60$.

\section{Summary and Conclusions}

We have described the dynamics of a BPS probe in the background
of a non-BPS brane. We have shown that inflation can only take
place after compactifying on a six dimensional manifold. The
resulting inflation occurs for any non-BPS brane. It ends well
before the final decay of the non-BPS brane; the only fine-tuning
being the choice of the initial positions of the branes which
determines the spectral index.

Of course the essential ingredient here is the lack of
supersymmetry which leads to a non-trivial potential for the
inter-brane distance. The actual decay of the non-BPS brane may
only play a role in the reheating phase where radiation starts
dominating due to the impinging radiation resulting from the
non-BPS brane annihilation. It would be very interesting to carry
out the same analysis for stable non-BPS brane  as
obtained in compact spaces directly. Of course one would have to
motivate the reheating phase differently.

As it stands the brane technology seems to lead to an almost unique
form of the inflationary potential, an inverse power law. This is
a tremendous improvement over the four dimensional approach where
all the potentials seem to be ad hoc and difficult to motivate.
On the other hand such a potential is reminiscent of the potentials
used to describe quintessence. It would be very interesting if stable
non-BPS configurations \cite{FR,MUI,MUII} could provide a unique description of both
inflation and quintessence. This is left for future work.

\acknowledgments

We thank Ruth Durrer and Jean-Philippe Uzan for useful
discussions.

\typeout{--- No new page for bibliography ---}

\end{document}